# Medication non adherence: finding solutions through design thinking approach

**Anna Iurchenko**
*Product Designer at Stanfy, San Francisco, USA*
aiurchenko@stanfy.com

**ABSTRACT**

Medical non-adherence increasingly is recognized as a major medical health problem. Approximately 50% of patients do not take their medications as prescribed and such poor adherence has been shown to result in complications, death, and increased health care costs [1, 2, 5].

This problem becomes even more significant for patients with chronic illness and those who need to take medications lifetime, like transplant patients. Studies show that one half of rejection episodes and 15% of graft losses happen due to immunosuppression medications non-adherence [3, 4].

This article explores factors that have an impact on non-compliant behavior among transplant patients: patient factors, illness factor, therapeutic regimen factors [5, 6, 7]. Using user-centered design thinking approach based on these analyses a set of the hypothesisis defined and discussed strategies to enhance adherence by using mobile technology and gamification techniques.

**KEYWORDS**

Transplant patients, mHealth; medical apps; medication non adherence; health behavior; eHealth; medical compliance, gamification.

## INTRODUCTION

According to a 2003 report published by the WHO, adherence rates in developed countries only about 50% [9].

Adherence is a key factor associated with the effectiveness of treatment and is particularly critical for medications prescribed for chronic conditions. Of all medication-related hospitalizations that occur in the United States, between 30% and 60% are the result [10] of poor medication adherence.

Studies show that 20% to 80% of patients make errors in taking medication and that 20% to 60% stop taking medications before being instructed to do so [11, 12]. With older populations, the literature concerning adherence reports that compliance rates range roughly from 38% to 57%, with an average rate of less than 45% [13, 14].

The most consistently reported factors impacting adherence are low literacy, lack of health insurance coverage, poor social support and family instability [15]. It has been reported that one-half of adults lack the skills to complete tasks related to following medication-label directions, comprehending nutrition labels, describing symptoms, or using a map to locate health facilities [15].

Decreased health literacy has been reported with increasing age, such that adults aged 60 years may not understand basic materials such as medication labels. Patient beliefs about the diagnosis and medication benefits also have been shown to affect the adherence rate regardless of whether the patient can afford the medication. This means that patients may have a low perceived need for the medication and more concerns about side effects, thus possibly choosing to forgo the cost of the medication based on their beliefs [16].

The following steps have been shown to increase adherence [15]: verifying the patient's understanding of the disease and its treatment and providing education where gaps in understanding exist; correlating medication-taking with other daily routines; using medication organizers or charts; reducing the pill burden, if possible; providing care support.

In this research the human-centered design process was used to find the innovative, impactful solution that would help to increase adherence.

Research methodology involved a user-centred design approach incorporating strategies of human factors engineering, cognitive behavior science and elements of social marketing, followed by an iterative prototype test phase within the target population. This research resulted in a concept of a mobile application that will create a positive reinforcement through the gamification experience. The concept applies to a wide range of patients but discussed in particular for transplant patients.

**METHOD**

User-Centered Design (UCD) is a multidisciplinary design approach based on the active involvement of users to improve the understanding of user and task requirements, and the iteration of design and evaluation. It is widely considered the key to product usefulness and usability - an effective approach to overcoming the limitations of traditional system-centered design.

The human-centered design process has three phases — the Inspiration phase, the Ideation phase, and the Implementation phase. By keeping people in the heart of the design process we are making sure that the final solution will be tailored to their needs.

In the Inspiration phase we learn directly from the people were are designing for and immersing ourselves in their lives and come to deeply understand their needs. In the Ideation phase we are making sense of what we learned, identifying opportunities for design, and prototype the solutions. And in the Implementation phase are bringing our solution to life, and eventually, to people.

In this study we focused on defining the solution to medical non adherence for patients who had transplant surgery on liver, kidney, heart, lung, intestine or pancreas.

For the research we selected transplant patients who had surgery within the last year and more than three years before. This allowed us to see more details on how transplant patient's life change with time and also it allowed us to explore either extreme of the spectrum. An idea that suits an extreme user will nearly certainly work for the majority too. More importantly, talking to extremes can spark creativity by exposing you to use cases that you'd never have imagined on your own.

Multiple studies were conducted - interviews, observations, diary studies, co-design sessions, surveys. The goal was to step into transplant patients' shoes, to understand their lives and solve problems from their perspectives.

**RESULTS**

Insights that were got from the users during the research phase and understanding of medical non-adherence due to previous researches [17] drove us to the several hypotheses with the focus on the mobile experience.

*1.1 Personalization based on the patients' needs*

Standard pills reminders products are extremely hard to use - they are generic and try to satisfy patients with the different condition by building monstrous interfaces. So the solution should be only about transplant patients and their special treatments.

*1.2 Smart guidance to the anti-rejection medicine*

During the first several months, patients feel overwhelmed with information; they need to learn about their new organ, how to take care of it, and how to avoid rejection and infection. So the solution should be a smart guide to the anti-rejection medicine.

We should be focusing on some of the most important medications and timing them appropriately — so users don't have to. Together with that user should have total control over the timing of the medication notifications.

*1.3 Positive reinforcement for medication adherence*

The patient's life is changing dramatically for the better within several weeks after the surgery and they often treat the day of the surgery as their second birthday. So visual language and style of the mobile solution should emphasize the new, energetic patient's life, far away from a hospital.

Patients rarely admit that they are missing medications. In fact, sometimes they even hardly notice that they miss their anti-rejection medicine - they don't feel the immediate impact but it can be very dangerous if users miss their drugs often.

Guilt, shame, or a sense of failure doesn't help here. So, instead of being strict and demanding the solution should be a friend. It shouldn't stress them or punish them. It's soft and gentle reminder.

*1.4 Engaging into the treatment through gamification techniques*

The reward system was created and tested with the target group of transplant patients [18].

By introducing to the patients "7 Day Challenge" where they needed to take their medications for 7 days in a row without missing a dose we aimed to shift the focus from "You have to take medications" to "I want to take my medications". After completing the challenge users are involved into the game that encourages them to earn rewards towards eventually gather them all.

| Activity | Points | Challenges completed |
|---|---|---|
| Take medications for 1 day | 1 | 0 |
| Take medications 3 days in a row | 3 | 0 |
| Take medications for 5 days | 5 | 0 |
| Take medications for 7 days in a row | 7 | 1 |
| Completed 3 challenges | 21 | 3 |
| Take medications 17 days in a row without omissions | 17 | 2 |
| Completed 5 challenges | 35 | 5 |
| Don't miss medications 4 weeks in a row | 28 | 4 |
| Completed 10 challenges | 70 | 10 |
| Completed 15 challenges | 105 | 15 |

Table 1. Awards system for the patient's activities

The developed gamification system for transplant patients was tested on the group of potential users and through iterations we got to the understanding of the game level system that will be impactful to the medication adherence (see Table 1).

## DISCUSSION AND CONCLUSIONS

All the hypothesis were combined to design a final mobile solution that was tested in the study [19] where sixty-seven renal transplant recipients were prospectively enrolled and randomized into 2 groups: 18 app users and 49 nonusers.

The variability in tacrolimus levels was calculated using the coefficient of variability (CV): CV = (standard deviation/mean tacrolimus) x 100. CV has been shown to be a critical indicator of chronic rejection determined during protocol biopsies of renal allografts.

In the patient cohort, CV was significantly lower in app users compared to nonusers at 1 month (27.7 vs 37.0, respectively, $P=0.014$). CV at 1 month was a significant predictor of app utilization (odds ratio 0.916; 95% confidence interval 0.858-0.977; $P=0.007$) [20].

The CV reduction in app users at 1 month has shown a potential of such mobile solution to improve medication adherence in the early postoperative period when patients may be at highest risk for medication non adherence.

The rate of missed medications was also lower by 28% among app users who completed at least three "7 Day Challenges" and we found the direct correlation between missed doses and level achieved in the game system. This shows a positive influence of the game element of the proposed solution to the medical adherence of the transplant patients.

Further investigation can explore what elements of the game design may be improved to engage more users and also establish more factors that show correlation with patient behavior and health.